\documentclass[fleqn,usenatbib,onecolumn]{mnras}

\usepackage{newtxtext,newtxmath}

\usepackage[T1]{fontenc}

\DeclareRobustCommand{\VAN}[3]{#2}
\let\VANthebibliography\thebibliography
\def\thebibliography{\DeclareRobustCommand{\VAN}[3]{##3}\VANthebibliography}

\usepackage{graphicx}	
\usepackage{amsmath}	

\title[SSC in  Afterglow stratified medium]{Closure Relations of Synchrotron Self-Compton in  Afterglow stratified medium and Fermi-LAT Detected Gamma-Ray Bursts}

\author[Fraija et al.]{
Nissim Fraija,$^{1}$\thanks{E-mail: nifraija@astro.unam.mx}
Maria G. Dainotti,$^{2,3,4}$
B. Betancourt Kamenetskaia,$^{5,6}$
D. Levine$^{7}$ and
A. Galvan-Gamez$^{1}$
\\
$^{1}$Instituto de Astronom\' ia, Universidad Nacional Aut\'onoma de M\'exico, Circuito Exterior, C.U., A. Postal 70-264, 04510 M\'exico City, M\'exico\\
$^{2}$National Astronomical Observatory of Japan, 2-21-1 Osawa, Mitaka, Tokyo 181-8588, Japan\\
$^{3}$Space Science Institute, Boulder, CO, USA\\
$^{4}$The Graduate University for Advanced Studies, SOKENDAI, Shonankokusaimura, Hayama, Miura District, Kanagawa 240-0193, Japan\\
$^{5}$Technical University of Munich, TUM School of Natural Sciences, Physics Department, James-Franck-Str 1, 85748 Garching, Germany\\
$^{6}$Max-Planck-Institut f\"ur Physik (Werner-Heisenberg-Institut), F\"ohringer Ring 6, 80805 Munich, Germany\\
$^{7}$Department of Astronomy, University of Maryland, College Park, MD 20742, USA
}

\date{Accepted XXX. Received YYY; in original form ZZZ}

\pubyear{2015}

\begin{document}
\label{firstpage}
\pagerange{\pageref{firstpage}--\pageref{lastpage}}
\maketitle

\begin{abstract}
The Second Gamma-ray Burst Catalog (2FLGC) was announced by the Fermi Large Area Telescope (Fermi-LAT) Collaboration. It includes 29 bursts with photon energy higher than 10 GeV. Gamma-ray burst (GRB) afterglow observations have been adequately explained by the classic synchrotron forward-shock model, however, photon energies greater than 10 GeV from these transient events are challenging, if not impossible, to characterize using this afterglow model. Recently, the closure relations (CRs) of the synchrotron self-Compton (SSC) forward-shock model evolving in a stellar wind and homogeneous medium was presented to analyze the evolution of the spectral and temporal indexes of those bursts reported in 2FLGC. In this work,  we provide the CRs of the same afterglow model, but evolving in an intermediate density profile ($\propto {\rm r^{-k}}$) with ${\rm 0\leq k \leq2.5}$, taking into account the adiabatic/radiative regime and with/without energy injection for any value of the electron spectral index. The results show that the current model accounts for a considerable subset of GRBs that cannot be interpreted in either stellar-wind or homogeneous afterglow SSC model. The analysis indicates that the best-stratified scenario is most consistent with ${\rm k=0.5}$ for no-energy injection and ${\rm k=2.5}$ for energy injection.
\end{abstract}




\section{Introduction}

The most intense transients in the Universe are gamma-ray bursts (GRBs). These phenomena, which may last anywhere from milliseconds to a few of hours \citep{1993ApJ...413L.101K, 1999PhR...314..575P, 2015PhR...561....1K} and have a spectrum given by the empirical Band function \citep{1993ApJ...413..281B}, are frequently seen in the keV-MeV energy range. After the initial burst of gamma rays dies down, we observe a much longer-lasting phenomenon we call ``afterglow" in wavelengths ranging from radio to GeV. This phenomenon occurs when a relativistic outflow ejected from the central engine interacts with the surrounding environment and transfers a significant portion of its energy to the surrounding circumburst medium \citep{1997ApJ...476..232M, 1999PhR...314..575P}.   The synchrotron radiation, which ranges from radio to gamma rays \citep{1998ApJ...497L..17S,2009MNRAS.400L..75K, 2010MNRAS.409..226K,2013ApJ...763...71A, 2016ApJ...818..190F}, and the synchrotron self-Compton (SSC) process, which up-scattering synchrotron photon at very-high energies \citep[VHE $\geq$ 10 GeV;][]{2019ApJ...885...29F, 2019ApJ...883..162F,2021ApJ...918...12F, 2017ApJ...848...94F,2019arXiv191109862Z}, are the primary mechanisms by which the shock-accelerated electrons inside it are cooled down. The temporal ($\alpha$) and spectral ($\beta$) indexes of these radiation processes evolving in different environments (e.g., homogeneous or stellar wind) predict the ``closure relations" (CRs)  $F_\nu \propto t^{-\alpha}\nu^{-\beta}$, which are unique in each scenario and therefore used to describe the afterglow observations.\\  
Regarding the analysis of the high-energy emission we can date back the first emission even to the previous century.
The first claim of a VHE photon was associated with GRB 940217, detected by the Energetic Gamma-Ray Experiment Telescope \citep[EGRET;][]{1994Natur.372..652H}. In that same year, GRB 941017 also presented a late-time, spectrally-rising GeV component \citep{2003Natur.424..749G}. The Second Gamma-ray Burst Catalog (2FLGC) was provided by \cite{Ajello_2019}, and it includes data from the first 10 years of program (from 2008 to 2018 August 4). The collection includes a sample of GRBs with temporarily-extended emission described with a power-law (PL) and a broken power-law (BPL) function, and photon energies above a few GeV, as well as 169 GRBs with high-energy emission higher than $\ge100$ MeV.  Those bursts described with a BPL function showed a temporal break in the light curve around several hundreds of seconds.  The light curves of the bursts that might be represented with a BPL function showed a temporal break in time somewhere in the vicinity of many hundreds of seconds.   The evolution of the temporal and spectral indexes of the synchrotron afterglow model is usually invoked to interpreted the temporarily-extended emission \citep{1998ApJ...497L..17S,2009MNRAS.400L..75K,2010MNRAS.409..226K}.  The high-energy emission is most likely to be dominated by the forward shock, an idea that is motivated not just by the temporarily-extended LAT emission, but also by the delayed onset \citep[e.g.,][]{2009Sci...323.1688A} and the inconsistency of the GeV flux with a simple extrapolation from the sub-MeV energy range as observed in a significant fraction of bursts \citep[e.g.][]{2008ApJ...689.1150A, 2011MNRAS.416.3089B}. While it is expected that if the forward shock dominates the high-energy emission, CR would be present, not all LAT light curves satisfy this condition.

\cite{2019ApJ...883..134T} systematically analyzed the CRs in a sample of 59 chosen LAT-detected bursts, taking into account temporal ($\alpha_{\rm LAT}$) and spectral ($\beta_{\rm LAT}$) indices. While the typical synchrotron forward-shock emission accomplishes its goals effectively of explaining the spectral and temporal indices in most instances, they reported that a sizeable minority of bursts defy this explanation.   \cite{2019ApJ...883..134T} reported that a large fraction of bursts fulfills the CRs when the synchrotron scenario: i) lies in the slow-cooling regime ($\nu^{\rm syn}_{\rm m}<\nu_{\rm LAT}<\nu^{\rm syn}_{\rm c}$), ii) evolves  in the homogeneous medium, and iii) requires that magnetic microphysical parameter has an atypically small value of $\varepsilon_B<10^{-7}$. The spectral breaks $\nu^{\rm syn}_{\rm c}$ and $\nu^{\rm syn}_{\rm m}$ corresponds to the cooling and characteristic ones, respectively.   On the other hand, \cite{2010MNRAS.403..926G} analyzed the high-energy emission of 11 bursts (up until October of 2009) identified by the Fermi LAT. They concluded that the relativistic forward shock was in the fully radiative regime for a hard spectral index $p\sim 2$ because the LAT observations evolved as $\sim t^{-\frac{10}{7}}$ rather than $t^{-1}$ as predicted by the adiabatic synchrotron forward-shock model.  They reported that three of them displayed this phase, being a slow-cooling regime in homogeneous interstellar medium (ISM) the most favorable scenario.  Furthermore, \cite{2021ApJS..255...13D} looked into the cases  with a broken PL segment with a shallower slope than the usual decay, the so-called ``plateau" phase, in the X-ray and optical light curves and found three cases with known redshift among a sample of GRBs deteted by Fermi-LAT from August 2008 until May 2016. One of the possibilities for interpreting this phase is provided by the the energy injection scenario \citep{2005ApJ...635L.133B,  2005ApJ...630L.113K, 2006Sci...311.1127D, 2006ApJ...636L..29P, 2006MNRAS.370L..61P, 2005Sci...309.1833B, 2007ApJ...671.1903C, 2017MNRAS.464.4399D, 2019ApJ...872..118B}. They found that the slow-cooling regime evolving in ISM was the most favourable scenario.  Finally, \cite{2022ApJ...934..188F} derived the CRs of SSC forward-shock model evolving in the stellar wind and ISM. The authors showed that a significant fraction of GRBs could not be interpreted in the synchrotron scenario, and also that ISM is preferred without energy injection and the stellar wind  with energy injection.\\

CRs evolving in ISM ($k=0$) and stellar wind ($k=2$) have been widely used to test the evolution of LAT light curves. However, studies about the progenitor and modelling of the multiwavelength observations during the afterglow phase has suggested that a circumburst medium with an intermediate value of density profile between ${\rm k}=0$ and ${\rm k}=2$ is consistent in several GRBs \citep{Dainotti2023Galax..11...25D}. For instance,  \cite{2013ApJ...776..120Y} investigated the evolution of the emission of forward-reverse shocks propagating in a medium described by the previously mentioned PL distribution. They applied their model to 19 GRBs and found that the density profile index took values of $ 0.4 \leq {\rm k}\leq  1.4$, with a typical value of ${\rm k\sim1}$. This value was also obtained by \cite{2013ApJ...774...13L}, who analyzed a bigger sample of 146 GRBs. A  density-profile index with $k>2$ was derived by \cite{2008Sci...321..376K} after looking at the possibility that a black hole's accretion of a stellar envelope causes the plateau phase in X-ray light curves. Also in GRBs presenting radio afterglows a stratified emission can reconcile some of the observations \citep{Levine2022MNRAS.tmp.3552L}. \cite{2013ApJ...779L...1K} analyzed the multi-wavelength observations of GRB 130427A, one of the most powerful burst with a redshift $z<1$, and modeled the observations below GeV energy range with a synchrotron afterglow model in the adiabatic regime without energy injection for $p>2$. The authors found that an intermediate value of density profile between ${\rm k}=0$ and ${\rm k}=2$ was consistent with these observations.\\

\cite{2022ApJ...934..188F} presented the CRs for the stellar-wind and ISM in the forward-shock scenario of a SSC in stratified medium for $k=0$ and $k=2$. They considered the evolution in both the adiabatic and radiative regimes,  with and without energy injection into the blastwave. Here, we extend this approaches and take into account the  stratified environment with a density profile $\propto r^{-k}$ with ${\rm k}$ lying in the range of  $0\leq k < 3$. We apply the current model to investigate the evolution of the spectral and temporal indexes of those GRBs reported in 2FLGC.  This manuscript will be organized as follows: in Section \S\ref{sec2} we derive the CRs of SSC forward-shock model evolving in a stratified environment with $0\leq k < 3$. In Section \S\ref{sec3}, we show the analysis of Fermi-LAT data and the methodology. Section \S\ref{sec4} displays the discussion and finally, in Section \S\ref{sec5}, we conclude by summing up our findings and making some general observations.

\section{SSC Forward-shock model in stratified medium}
\label{sec2}
By decelerating and driving a forward shock into the circumstellar medium, a relativistic GRB outflow creates the temporally-extended afterglow emission. During the deceleration phase, the outflow transmits a large portion of its energy to the external medium. Shocks transmit a significant amount of energy density to accelerate electrons and increase the magnetic field through the microphysical parameters $\varepsilon_e$ and  $\varepsilon_B$, respectively.   We take into account a density-profile described by $n(r) =A_{\rm k}\, r^{\rm -k}$, with $0\leq k < 3$ in which accelerated electrons cool down.  Based on the foregoing, we derive the CRs of SSC afterglow in adiabatic/radiative regime and with/without energy injection. In order to report the proportionality constant of each quantity, we consider the values of PL indexes $k=1.0$ for $A_{\rm k}=1.52\times 10^{15}\,{\rm cm^{-2}}$, $\epsilon=0.5$, $q=0.5$, and $p=1.95$ ($2.15$) for $1<p<2$ ($2\leq p$), unless otherwise stated.

\subsection{Closure relations in the adiabatic and radiative scenario}

The relativistic forward shock is often assumed to be totally adiabatic in the conventional GRB afterglow scenario, however it is possible that it is either partially or completely radiative \citep{1998MNRAS.298...87D, 1998ApJ...497L..17S, 2000ApJ...532..281B, 2010MNRAS.403..926G}.  Comparing the cooling and dynamical timescales,  the relativistic electrons could lie in the fast or slow-cooling regime.   In the case of the cooling timescale being smaller than the dynamical timescale, the relativistic electrons are in the fast-cooling regime, and only when $\varepsilon_e$ to be of order unity,  the afterglow phase would be in the radiative regime \citep{2000ApJ...532..281B,2002MNRAS.330..955L,  2019ApJ...886..106P}.  The deceleration of the GRB fireball by the circumburst medium is faster when the fireball is radiative than adiabatic. As a result, the shock's energetics and  synchrotron light curves during the afterglow are altered \citep{2000ApJ...532..281B, 2005ApJ...619..968W, 2000ApJ...529..151M}.\\

Once the relativistic jet begins the deceleration phase, the evolution of the bulk Lorentz factor first in the radiative and then in the adiabatic regime is $\Gamma=2.8\times 10^2 \left(\frac{1+z}{1.022}\right)^{\frac{3-k}{[2(4-k)-\epsilon]}}\, A_k^{-\frac{1}{[2(4-k)-\epsilon]}}E_{53}^{\frac{1}{[2(4-k)-\epsilon]}}\Gamma_{0,2.8}^{-\frac{\epsilon}{[2(4-k)-\epsilon]}}t_{2}^{-\frac{3-k}{[2(4-k)-\epsilon]}}$ \citep{2005ApJ...619..968W}, where $E$ is the isotropic-equivalent kinetic energy, and $\Gamma_0$ is the initial Lorentz factor.  Hydrodynamic evolution of the forward shock is accounted for in the totally adiabatic ($\epsilon=0$), fully radiative ($\epsilon=1$), or partly radiative or adiabatic ($0 <\epsilon < 1$) regimes, respectively, where $\epsilon$ is the radiative parameter. Hereafter, we adopt the convention $Q_{\rm x}=Q/10^{\rm x}$ in c.g.s. units.   For electrons with the lowest energy, above which they cool effectively and with the maximum energy, the Lorentz factors for $p>2$ are

{\small
\begin{eqnarray}
\label{ele_Lorent_ism}
\gamma_{\rm m}&=& 8.6\times 10^3\,\left(\frac{p-2}{p-1}\right) \left(\frac{1+z}{1.022}\right)^{\frac{3-k}{[2(4-k)-\epsilon]}}\, \varepsilon_{e,-0.3} A_k^{-\frac{1}{[2(4-k)-\epsilon]}}\, E_{53}^{\frac{1}{[2(4-k)-\epsilon]}}\, \Gamma^{-\frac{\epsilon}{[2(4-k)-\epsilon]}}_{0,2.8}   t_{2}^{-\frac{3-k}{[2(4-k)-\epsilon]}} \cr
\gamma_{\rm c}&=& 1.9\times 10^{4}\,(1+Y)^{-1} \left(\frac{1+z}{1.022}\right)^{-\frac{1+\epsilon+k(1-\epsilon)}{[2(4-k)-\epsilon]}}\, \varepsilon^{-1}_{B,-2} A_k^{-\frac{5-\epsilon}{[2(4-k)-\epsilon]}}\, E_{53}^{-\frac{3-2k}{[2(4-k)-\epsilon]}}\, \Gamma^{\frac{(3-2k)\epsilon}{[2(4-k)-\epsilon]}}_{0,2.8}   t_{2}^{\frac{1+\epsilon+k(1-\epsilon)}{[2(4-k)-\epsilon]}}\,\cr
\gamma_{\rm max}&=& 5.9\times 10^{7}\, \left(\frac{1+z}{1.022}\right)^{-\frac{6-k\epsilon}{4[2(4-k)-\epsilon]}}\, \varepsilon^{-\frac14}_{B,-2} A_k^{-\frac{6-\epsilon}{4[2(4-k)-\epsilon]}}\, E_{53}^{-\frac{1-k}{2[2(4-k)-\epsilon]}}\, \Gamma^{\frac{(1-k)\epsilon}{2[2(4-k)-\epsilon]}}_{0,2.8}   t_{2}^{\frac{6-k\epsilon}{4[2(4-k)-\epsilon]}}\,,\hspace{6.cm}
\end{eqnarray}
}

respectively, where  $Y$ is the Compton parameter and $t$ is the observer time.  The Lorentz factor of the lowest-energy electrons for ${\rm 1<p<2}$ is 

{\small
\begin{equation}\label{nus_ism}
\gamma_{\rm m}= 2.1\times 10^3\,\left(\frac{2-p}{p-1}\right)^{\frac{1}{p-1}}\left(\frac{1+z}{1.022}\right)^{\frac{6(4-p)-k[4+(2-p)\epsilon]}{4(p-1)[2(4-k)-\epsilon]}}\, \varepsilon^{\frac{1}{p-1}}_{e,-0.3}  \varepsilon^{\frac{2-p}{4(p-1)}}_{B,-2}  A_k^{\frac{2(4-3p)+\epsilon(p-2)}{4(p-1)[2(4-k)-\epsilon]}}E_{53}^{\frac{4-p-k(2-p)}{2(p-1)[2(4-k)-\epsilon]}}\, \Gamma^{-\frac{[4-p-k(2-p)]}{2(p-1)[2(4-k)-\epsilon]}}_{0,2.8}   t_{2}^{-\frac{6(4-p)-k[4+(2-p)\epsilon]}{4(p-1)[2(4-k)-\epsilon]}}\,,  
\end{equation}
}

In the synchrotron afterglow model, the characteristic and cooling spectral breaks evolve as $\nu^{\rm syn}_{\rm m}\propto t^{-\frac{6(p+2)-k(2+2p+\epsilon)}{2(p-1)[2(4-k)-\epsilon]}}$ ($t^{-\frac{24-k(6+\epsilon)}{2[2(4-k)-\epsilon]}}$) for $1<p<2$ ($2<p$) and $\nu^{\rm syn}_{\rm c}\propto t^{\frac{(\epsilon-2)(4-3k)}{2[2(4-k)-\epsilon]}}$, respectively, and the maximum synchrotron flux as $F^{\rm syn}_{\rm max}\propto t^{-\frac{k(2-3\epsilon)+6\epsilon}{2[2(4-k)-\epsilon]}}$. When relativistic electrons are accelerated in stratified environments in the radiative regime,  synchrotron photons have their maximum energy correspond to 

\begin{equation}
\label{ene_max_eps}
h\nu^{\rm syn}_{\rm max} \approx 
5.4\times 10^2\, {\rm MeV}\left(\frac{1+z}{1.022}\right)^{\frac{k+\epsilon-5}{2(4-k)-\epsilon}}\,  A_k^{-\frac{1}{2(4-k)-\epsilon}} E_{53}^{\frac{1}{2(4-k)-\epsilon}}\Gamma_{0,2.8}^{-\frac{\epsilon}{2(4-k)-\epsilon}}t_{2}^{-\frac{3-k}{2(4-k)-\epsilon}}\,.
\end{equation}

When the same population of electrons that emits synchrotron photons also up-scatters them to higher energies, a phenomenon known as  SSC emission occurs; $h\nu^{\rm ssc}_{\rm i}\simeq \gamma^2_{\rm i} h\nu^{\rm syn}_{\rm i}$, with ${\rm i=m, c}$ \citep[e.g., see][]{2001ApJ...548..787S}. The maximum flux generated by the SSC mechanism is $F^{\rm ssc}_{\rm max}\sim \sigma_T n\,r\, F^{\rm syn}_{\rm max}$.\footnote{$\sigma_T$ is the Thompson cross section and $r$ is the forward-shock radius.}  Requiring the Lorentz factors for electrons (Eq. \ref{ele_Lorent_ism}), and the relevant quantities (spectral breaks and the maximum flux) for the synchrotron scenario,  the characteristic (for $2 < p$) and cooling spectral breaks and the maximum flux of the SSC scenario  are

{\small
\begin{eqnarray}
\label{break_ssc_hom}
h\nu^{\rm ssc}_{\rm m} &=&
1.7\times 10^9\,{\rm eV}\,\left(\frac{p-2}{p-1}\right)^4 \left(\frac{1+z}{1.022}\right)^{\frac{2(10+\epsilon)-k(6+\epsilon)}{2[2(4-k)-\epsilon]}}\, \varepsilon^4_{e,-0.3} \varepsilon^{\frac12}_{B,-2} 
A_k^{-\frac{4+\epsilon}{2[2(4-k)-\epsilon]}}\, E_{53}^{\frac{6-k}{[2(4-k)-\epsilon]}}\, \Gamma^{-\frac{(6-k)\epsilon}{[2(4-k)-\epsilon]}}_{0,2.8}   t_{2}^{-\frac{36-k(10+\epsilon)}{2[2(4-k)-\epsilon]}} \cr
h\nu^{\rm ssc}_{\rm c}&=& 4.5\times 10^{10}\,{\rm eV} \left(1+Y\right)^{-4} \left(\frac{1+z}{1.022}\right)^{-\frac{6(2+\epsilon)+k(6-7\epsilon)}{2[2(4-k)-\epsilon]}}\, \varepsilon^{-\frac72}_{B,-2} A_k^{-\frac{36-7\epsilon}{2[2(4-k)-\epsilon]}}\, E_{53}^{-\frac{10-7k}{[2(4-k)-\epsilon]}}\, \Gamma^{\frac{(10-7k)\epsilon}{[2(4-k)-\epsilon]}}_{0,2.8} \,t_{2}^{-\frac{4-8\epsilon-k(10-7\epsilon)}{2[2(4-k)-\epsilon]}}\cr
F^{\rm ssc}_{\rm max}&=& 1.0\times 10^{-8}\,{\rm mJy}\left(\frac{p-2}{p-1}\right)^{-1} \left(\frac{1+z}{1.022}\right)^{\frac{4(7+\epsilon)-k(2+5\epsilon)}{2[2(4-k)-\epsilon]}}\,  \varepsilon^\frac12_{B,-2} D_{\rm z,26.5}^{-2}\,A_k^{\frac{5(4-\epsilon)}{2[2(4-k)-\epsilon]}}\, E_{53}^{\frac{5(2-k)}{[2(4-k)-\epsilon]}}\, \Gamma^{-\frac{5(2-k)\epsilon}{[2(4-k)-\epsilon]}}_{0,2.8}   t_{2}^{\frac{4-k(6-5\epsilon)-8\epsilon}{2[2(4-k)-\epsilon]}}\,,
\end{eqnarray}
}

respectively, and the characteristic break for ${1<p<2 }$ is

{\small
\begin{equation} \label{ssc_br_hom_v1}
h\nu^{\rm ssc}_{\rm m}=5.7\times 10^6\,{\rm eV}\,\left(\frac{2-p}{p-1}\right)^{\frac{4}{p-1}} \left(\frac{1+z}{1.022}\right)^{\frac{52-2p(8-\epsilon)-2\epsilon-k[10+3\epsilon-p(2+\epsilon)]}{2(p-1)[2(4-k)-\epsilon]}}\, \varepsilon^{\frac{4}{p-1}}_{e,-0.3}  \varepsilon^{\frac{3-p}{2(p-1)}}_{B,-2}  A_k^{\frac{3(4-\epsilon)-p(8-\epsilon)}{2(p-1)[2(4-k)-\epsilon]}}E_{53}^{\frac{6-k(3-p)}{(p-1)[2(4-k)-\epsilon]}}\, \Gamma^{-\frac{[6-k(3-p)]\epsilon}{(p-1)[2(4-k)-\epsilon]}}_{0,2.8}   t_{2}^{-\frac{36-kp(2-\epsilon)-3k(2+\epsilon)}{2(p-1)[2(4-k)-\epsilon]}}.
\end {equation}
}
When the Klein - Nishina (KN) effects are not negligible, the value of the Compton parameter is not constant and then defined as {\small $\frac{Y(\gamma_c)[Y(\gamma_c)+1]}{Y_0} =\left( \frac{\nu^{\rm syn}_{\rm m}}{\nu^{\rm syn}_{\rm c}} \right)^{-\frac{p-3}{2}}\,\left( \frac{\nu^{\rm syn}_{\rm KN}(\gamma_{\rm c})}{\nu^{\rm syn}_{\rm m}}\right)^\frac43 \,\,{\rm for}\,\,{\nu^{\rm syn}_{\rm KN}(\gamma_{\rm c}) <\nu^{\rm syn}_{\rm m} }$},  {\small $ \left(\frac{\nu^{\rm syn}_{\rm KN}(\gamma_{\rm c})}{\nu^{\rm syn}_c}\right)^{-\frac{p-3}{2}}\,\,{\rm for}\,\,    \nu^{\rm syn}_{\rm m} < \nu^{\rm syn}_{\rm KN}(\gamma_{\rm c}) < \nu^{\rm syn}_{\rm c}$},  {\small $1\,\, {\rm for} \,\, \nu^{\rm syn}_{\rm c} < \nu^{\rm syn}_{\rm KN}(\gamma_{\rm c})$} where {\small $Y_0=\frac{\varepsilon_{e}}{\varepsilon_{B}}  \left(\frac{\gamma_{\rm c}}{\gamma_{\rm m}}\right)^{2-p}$}  where $h\nu^{\rm syn}_{\rm KN} (\gamma_c) \simeq\frac{\Gamma}{(1+z)}\,\frac{m_e c^2}{\gamma_c}$ for $\nu^{\rm syn}_{\rm m} < \nu^{\rm syn}_{\rm c}$ \citep[see][]{2009ApJ...703..675N, 2010ApJ...712.1232W}.   To describe the LAT observations above 100 MeV, it is needed to define the Lorentz factor of those electrons ($\gamma_{*}$) that might emit high-energy photons via synchrotron process, and a new spectral break $h\nu^{\rm syn}_{\rm KN}(\gamma_*)$ would have to be included and the Compton parameter $Y(\gamma_{*})$ re-computed.  For instance, the new Compton parameter becomes {\small $Y(\gamma_*)=Y(\gamma_c) \left(\frac{\nu_{*}}{\nu_c}\right)^{\frac{p-3}{4}}   \left(\frac{\nu^{\rm syn}_{\rm KN}(\gamma_{\rm c})}{\nu^{\rm syn}_c}\right)^{-\frac{p-3}{2}}$} for $ \nu^{\rm syn}_{\rm m} <  h\nu^{\rm syn}_{\rm KN}(\gamma_*)=100\,{\rm MeV} < \nu^{\rm syn}_{\rm c} < \nu^{\rm syn}_{\rm KN}(\gamma_{\rm c})$  \citep[for details, see][]{2010ApJ...712.1232W}.

It is worth noting that the spectral breaks and the maximum flux derived for $k=0$ and $k=2$ and presented in \cite{2022ApJ...934..188F} are recovered.   Taking into account the temporal and spectral evolution of spectral breaks together with the maximum flux of the SSC afterglow model (Eqs. \ref{ssc_br_hom_v1}), we report the CRs as shown in Table \ref{Table1}.

\subsubsection{Evolution of the radiative parameter}

The evolution of radiative parameter can be estimated from the relation of synchrotron and expansion timescale \citep{2000ApJ...543...90H}. Therefore, it becomes

\begin{equation} \label{eps_t}
    \epsilon = \varepsilon_{e}\frac{{t'}^{-1}_{\rm syn}}{{t'}^{-1}_{\rm syn}+{t'}^{-1}_{\rm ex}}\,,
\end{equation}

where $t'_{\rm syn}= 6\pi m_e/(\sigma_T {B'}^2\,(1+Y(\gamma_{\rm c})) \gamma_m)$ is the synchrotron cooling time and $t'_{\rm exp}=4\,\Gamma t/(1+z)$ is the comoving frame expansion time  \citep[e.g., see][]{1998MNRAS.298...87D}, with $B'$ the comoving magnetic field, $m_e$ the electron mass and $\gamma_m$ given in Eqs. \ref{ele_Lorent_ism} and \ref{nus_ism} for $2\leq p$ and $1<p<2$, respectively. In order to discuss the evolution of the radiative parameter, we consider the limiting cases of synchrotron and expansion timescale.  For $t'_{\rm syn}\ll t'_{\rm exp}$, the radiative parameter corressponds to  $\epsilon = \varepsilon_{e}(1+t'_{\rm syn}/t'_{\rm exp})^{-1}\approx \varepsilon_{e}$ (fully radiative with $\varepsilon_{e}\approx 1$) and for $t'_{\rm exp}\ll t'_{\rm syn}$, the radiative parameter becomes $\epsilon = \varepsilon_{e}\, t'_{\rm exp}/t'_{\rm syn} (1+t'_{\rm exp}/t'_{\rm syn})^{-1}\approx 0$ (fully adiabatic). For instance, the radiative parameter becomes $\epsilon \approx \varepsilon_{e}/2$ when $t'_{\rm exp}\approx t'_{\rm syn}$.

\subsection{Closure relations with energy injection}\label{sec_22}

Refreshed shocks can be generated by the continuous injection of energy by the progenitor into the afterglow.  An expression for the luminosity injected by the progenitor $L_{\rm inj}(t)\propto L_0\, t^{\rm -q}$  where $q$ is the energy injection index \citep[e.g.][]{2006ApJ...642..354Z}. The equivalent kinetic energy of the forward shock is given by $E =\int L_{\rm inj}\, dt\propto L_0 t^{-q+1}$.    The temporal evolution of the bulk Lorentz factor can be written as  {\small $\Gamma=3.3\times 10^2\left(\frac{1+z}{1.022}\right)^{\frac{3-k}{2(4-k)}}\,  A_k^{-\frac{1}{2(4-k)}}E_{53}^{\frac{1}{2(4-k)}} t_{2}^{-\frac{2+q-k}{2(4-k)}}$}.

For electrons with the lowest energy, above which they cool effectively and with the maximum energy, the Lorentz factors for $p>2$ are

{\small
\begin{eqnarray}\label{gam_m_i}  
\gamma_{\rm m}&=& 1.0\times 10^4\,\left(\frac{p-2}{p-1}\right)\left(\frac{1+z}{1.022}\right)^{\frac{3-k}{2(4-k)}}\, \varepsilon_{e,-1} A_k^{-\frac{1}{2(4-k)}}\, E_{53}^{\frac{1}{2(4-k)}}\,  t_{2}^{-\frac{2+q-k}{2(4-k)}}\cr
\gamma_{\rm c}&=& 1.7\times 10^4\,(1+Y)^{-1} \left(\frac{1+z}{1.022}\right)^{-\frac{1+k}{2(4-k)}}\, \varepsilon^{-1}_{B,-2} A_k^{-\frac{5}{2(4-k)}}\, E_{53}^{-\frac{3-2k}{2(4-k)}}\,  t_{2}^{-\frac{2-3q-k(3-2q)}{2(4-k)}} \cr
\gamma_{\rm max}&=& 5.9\times 10^7\,\left(\frac{1+z}{1.022}\right)^{-\frac{3}{4(4-k)}}\,  \varepsilon^{-\frac14}_{B,-2}\,n^{-\frac{3}{4(4-k)}}\, E^{-\frac{1-k}{4(4-k)}}_{53}\,  t_{2}^{\frac{2+q+k(1-q)}{4(4-k)}}\,,\hspace{7cm}
\end{eqnarray}
}

respectively.   The Lorentz factor of the lowest-energy electrons for ${\rm 1<p<2}$ is 

{\small
\begin{equation}\label{nu_c_i}
\gamma_{\rm m}= 2.4\times 10^3\,\left(\frac{2-p}{p-1}\right)^\frac{1}{p-1} \left(\frac{1+z}{1.022}\right)^{\frac{3(4-p)-2k}{4(p-1)(4-k)}}\, \varepsilon^{\frac{1}{p-1}}_{e,-1}  \varepsilon^{\frac{2-p}{4(p-1)}}_{B,-2}  A_k^{\frac{4-3p}{4(p-1)(4-k)}} E_{53}^{\frac{4-k(2-p)-p}{4(p-1)(4-k)}}\,t_{2}^{-\frac{8+2(2-k)q-p(2+k+q-kq)}{4(p-1)(4-k)}}\,,
\end{equation}
}

respectively.   In the synchrotron afterglow model, the characteristic and cooling spectral breaks evolve as $\nu^{\rm syn}_{\rm m}\propto t^{-\frac{4+(2-k)q+p(2-k+q)}{2(p-1)(4-k)}}$ ($\nu^{\rm syn}_{\rm m}\propto t^{-\frac{2+q}{2}}$) for $1<p<2$ ($2<p$) and $\nu^{\rm syn}_{\rm c}\propto t^{\frac{(4-3k)(q-2)}{2(4-k)}}$, respectively, and the maximum synchrotron flux as $F^{\rm syn}_{\rm max}\propto t^{\frac{8(1-q)-k(4-3q)}{2(4-k)}}$.  For the adiabatic regimes and with energy injection, the maximum energy of synchrotron photons yields 

\begin{equation}
\label{ene_max_q}
h\nu^{\rm syn}_{\rm max} \approx
6.4\times 10^2\, {\rm MeV}\left(\frac{1+z}{1.022}\right)^{\frac{k-5}{2(4-k)}}\,    A_k^{-\frac{1}{2(4-k)}}E_{53}^{\frac{1}{2(4-k)}}\,t_{2}^{-\frac{2+q-k}{2(4-k)}}\,.
\end{equation}

Requiring the Lorentz factors for electrons (Eq. \ref{ele_Lorent_ism}), and the relevant quantities (spectral breaks and the maximum flux) for the synchrotron scenario,  the characteristic (for $2 < p$) and cooling spectral breaks and the maximum flux of the SSC scenario  are

{\small
\begin{eqnarray}
\label{syn_esp_win}
h\nu^{\rm ssc}_{\rm m}&=&  3.7\times 10^9\,{\rm eV}\left(\frac{p-2}{p-1}\right)^4\left(\frac{1+z}{1.022}\right)^{\frac{10-3k}{2(4-k)}}\, \varepsilon^4_{e,-1} \varepsilon^{\frac12}_{B,-2} A_k^{-\frac{1}{4-k}}\, E^{\frac{6-k}{2(4-k)}}_{53}\,   t_{2}^{-\frac{6(2+q)-k(4+q)}{2(4-k)}}\cr
h\nu^{\rm ssc}_{\rm c}&=& 2.9\times 10\,{\rm eV} (1+Y)^{-4} \left(\frac{1+z}{1.022}\right)^{-\frac{3(2+k)}{2(4-k)}}\, \varepsilon^{-\frac72}_{B,-2} A_k^{-\frac{9}{4-k}}\, E_{53}^{-\frac{10-7k}{2(4-k)}}\, t_{2}^{-\frac{12-10q-k(12-7q)}{2(4-k)}}\cr
F^{\rm ssc}_{\rm max}&=& 2.2\times 10^{-8}\,{\rm mJy}\left(\frac{p-2}{p-1}\right)^{-1} \left(\frac{1+z}{1.022}\right)^{\frac{14-k}{2(4-k)}}\,  \varepsilon^\frac12_{B,-2} D_{\rm z,26.5}^{-2}\,A_k^{\frac{5}{4-k}}\, E_{53}^{\frac{5(2-k)}{2(4-k)}}\, t_{2}^{\frac{2(6-5q)-k(8-5q)}{2(4-k)}}\,,
\end{eqnarray}
}
respectively, and the characteristic break for ${ 1<p<2 }$ is

{\small
\begin{equation} \label{ssc_br_hom_q}
h\nu^{\rm ssc}_{\rm m} =
1.3\times 10^{7}\,{\rm eV} \left(\frac{2-p}{p-1}\right)^{\frac{4}{p-1}} \left(\frac{1+z}{1.022}\right)^{\frac{26-k(5-p)-8p}{2(p-1)(4-k)}}\, \varepsilon^{\frac{4}{p-1}}_{e,-1}  \varepsilon^{\frac{3-p}{2(p-1)}}_{B,-2}  A_k^{\frac{3 - 2p }{(p-1)(4-k)}}  E^{\frac{6-k(3-p)}{2(p-1)(4-k)}}_{53}\,t_{2}^{-\frac{6(2+q)-3kq-kp(2-q)}{2(p-1)(4-k)}}
\end{equation}
}

It is important to point out that the spectral breaks and the maximum flux derived for $k=0$ and $k=2$ and presented in \cite{2022ApJ...934..188F} are recovered.

Taking into account the temporal and spectral evolution of spectral breaks together with the maximum flux of the SSC afterglow model (Eqs. \ref{ssc_br_hom_q}), we report the CRs as shown in Table \ref{Table2}.

\section{Methods and Analysis}
\label{sec3}

\begin{figure*} 
\centering
\begin{tabular}{ccc}
    \includegraphics[width=0.94\textwidth]{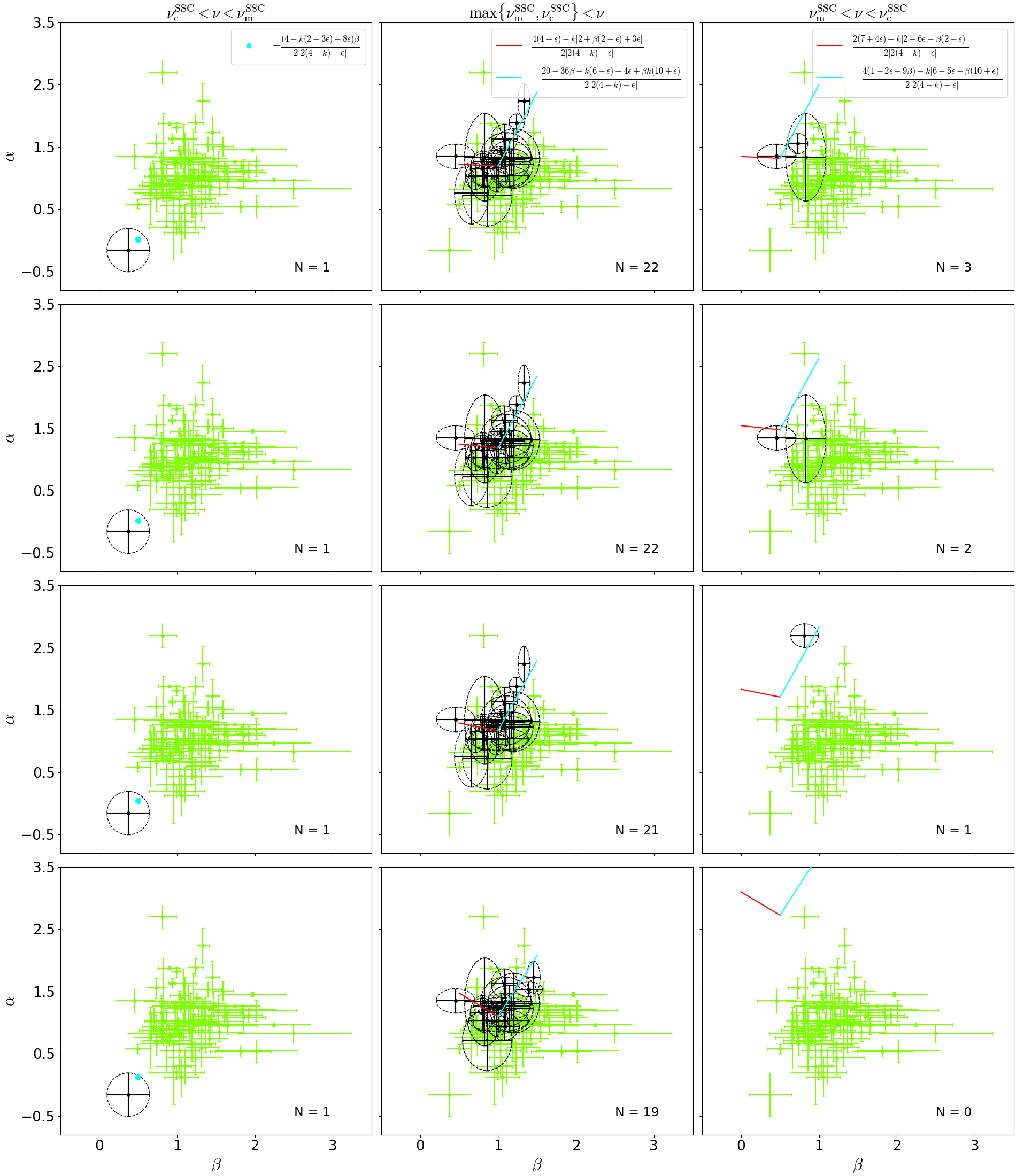} \\
\end{tabular}
\caption{The panels display the closure relations of SSC model in fast and slow cooling regimes for $\epsilon=0.5$ without energy injection with the spectral and temporal indexes from 2FLGC.  Panels from top to bottom are $k=0.5$, $1$, $1.5$ and $2.5$, respectively. The purple ellipses are shown when the CR are satisfied within 1 sigma error bars.} \label{Fig1}
\end{figure*}

\begin{figure*} 
\centering
\begin{tabular}{ccc}
    \includegraphics[width=0.94\textwidth]{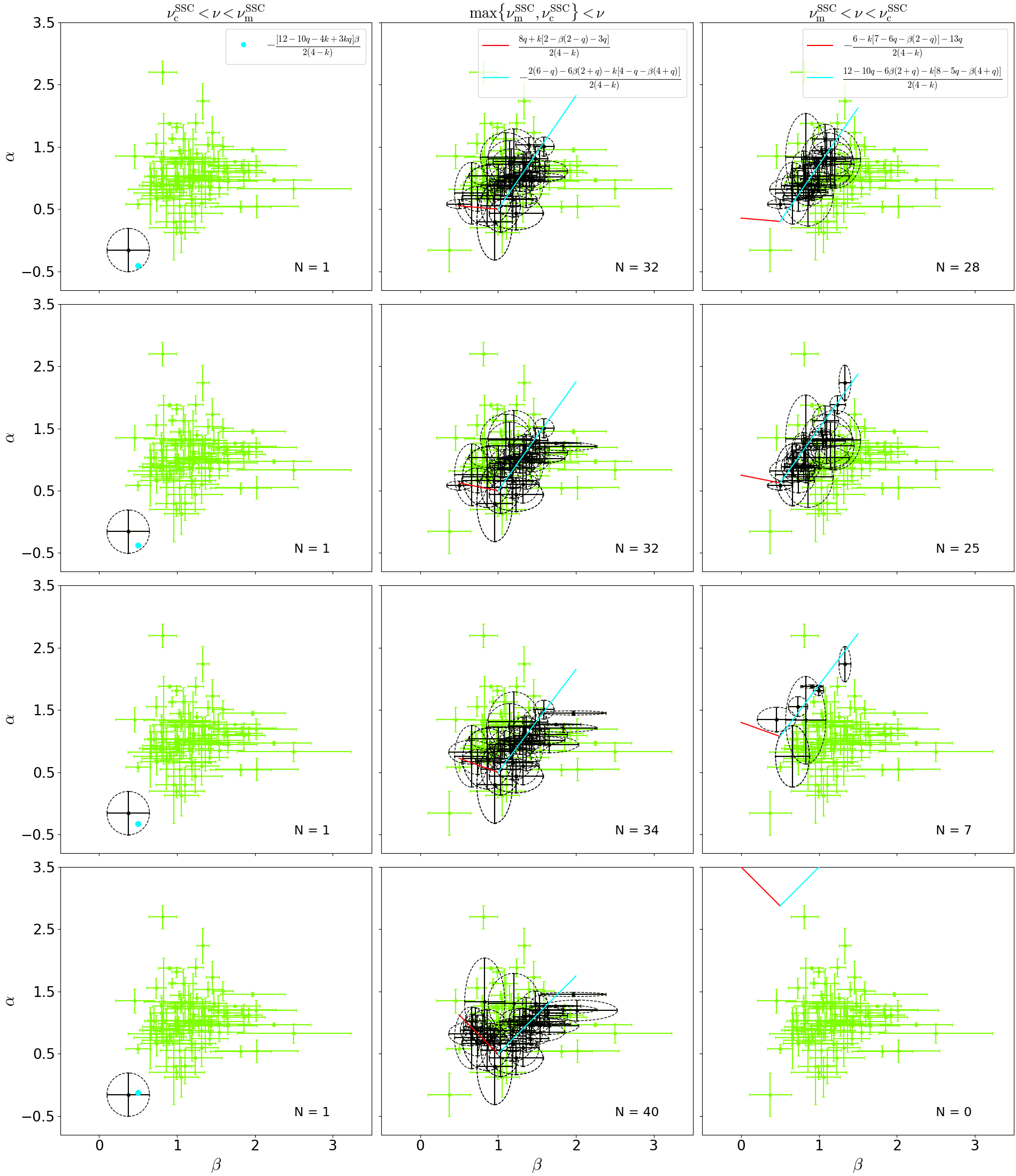} \\
\end{tabular}
\caption{The same as Fig. \ref{Fig1}, but with energy injection ($q=0.5$) and in the adiabatic regime.} \label{Fig2}
\end{figure*}

Our sample consists of 86 LAT-detected bursts with temporally extended emissions considered in \cite{2022ApJ...934..188F} and reported by the 2FLGC \citep{Ajello_2019}.  We analyze the evolution of temporal and spectral indexes using the same strategy presented in \cite{2022ApJ...934..188F,2021PASJ...73..970D, 2021ApJS..255...13D, 2020ApJ...903...18S}. The errorbars in the plots are in the 1$\sigma$ range, with $\alpha$ and $\beta$ values of the PL segment $F_{\rm \nu} \propto t^{-\alpha} \nu^{-\beta}$ being dependent; thus, Figures \ref{Fig1} and \ref{Fig2} are displayed with ellipses. 

We have derived and listed in Tables \ref{Table1} and \ref{Table2} the CRs of SSC afterglow model in the adiabatic and radiative scenario, and with and without energy injection, respectively.  We consider the CRs as function of the density-profile index $k$, the radiative parameter $\epsilon$, the energy injection index $q$, and the electron spectral index for $1<p<2$ and $ 2\leq p$. We emphasize that although early analysis of LAT data (using lower energy photons) found CRs that were consistent with the synchrotron FS model for the cooling condition $\nu^{\rm syn}_{\rm c}< \nu$ \citep[e.g.][]{2010MNRAS.409..226K}, we only consider the CRs of the SSC FS model with a sample of 86 LAT-detected bursts with temporally extended emissions ($\geq 100\,{\rm MeV}$) reported by the 2FLGC \citep{Ajello_2019}. Tables \ref{Table3} and \ref{Table4} summarize the number and percentage of bursts following each scenario.

Figure \ref{Fig1} presents the radiative case ($\epsilon=0.5$) in which no energy is injected. This figure exhibits the CR of SSC afterglow model evolving in a stratified medium, from top to bottom $k=0.5$, $1$, $1.5$ and $2.5$, respectively. The fast and slow regimes for each cooling condition are shown for $1<p<2$ and $p>2$. Table \ref{Table3} summarizes the number and percentage of bursts following each cooling condition for $1<p<2$ and $p>2$ and  the profile index $k=0.5$, $1$, $1.5$ and $2.5$.  This best-case scenario is most consistent with ${\rm max \{\nu^{\rm ssc}_{\rm m}, \nu^{\rm ssc}_{\rm c}\}<\nu}$ for $k=0.5$ (22 GRBs, 25.88\%), $k=1.0$ (22 GRBs, 25.88\%), $k=1.5$ (21 GRBs, 24.71\%) and $k=2.5$ (19 GRBs, 22.35\%).  There is one burst that satisfies the CRs for each value of $k$ when the SSC model evolve in the cooling condition  ${\rm \nu^{\rm ssc}_{\rm c}\leq \nu \leq  \nu^{\rm ssc}_{\rm m}}$, and few bursts for  ${\rm \nu^{\rm ssc}_{\rm m}\leq \nu \leq  \nu^{\rm ssc}_{\rm c}}$ for $k=0.5$ (3 GRBs, 3.53\%), $1$ (2 GRBs, 2.35\%) and $1.5$ (1 GRB, 1.18\%).\\

Figure \ref{Fig2} presents the adiabatic case in which energy ($q=0.5$) is injected. This Figure exhibits the CR of SSC afterglow model evolving in a stratified medium, from top to bottom $k=0.5$, $1$, $1.5$ and $2.5$, respectively. The fast and slow regimes for each cooling condition are shown for $1<p<2$ and $p>2$. Table \ref{Table4} summarizes the number and percentage of bursts following each cooling condition for $1<p<2$ and $p>2$ and  the profile index $k=0.5$, $1$, $1.5$ and $2.5$. The best-case scenario is most consistent with ${\rm max \{\nu^{\rm ssc}_{\rm m}, \nu^{\rm ssc}_{\rm c}\}<\nu}$ for $k=0.5$ (32 GRBs, 37.65\%), $k=1.0$ (32 GRBs, 37.65\%), $k=1.5$ (34 GRBs, 40.0\%) and $k=2.5$ (40 GRBs, 47.06\%).  For the cooling condition ${\rm \nu^{\rm ssc}_{\rm m}\leq \nu \leq  \nu^{\rm ssc}_{\rm c}}$, 28 (25 and 7) GRBs for $k=0.5$ ($1$ and $1.5$) fulfill the CRs.  There is one burst that satisfies the CRs for each value of $k$ when the SSC model evolve in the cooling condition  ${\rm \nu^{\rm ssc}_{\rm c}\leq \nu \leq  \nu^{\rm ssc}_{\rm m}}$.


\section{Discussion}
\label{sec4}

\subsection{A particular case: GRB 130427A}



The long GRB 130427A has been one of the most energetic bursts reported in the second LAT burst catalog. \cite{2013ApJ...779L...1K} performed an analysis on the multi-wavelength data of GRB 130427A and modeled the observations in the sub-GeV energy range using a synchrotron afterglow model operating in the adiabatic domain and without energy injection for p values greater than 2. They showed that an intermediate value of the density profile for constant ($\rm k=0$) and stellar ($\rm k=2$) medium was compatible with these results. In addition, they hypothesized that the enormous progenitor may have erupted before the core collapse event, which would have caused a change in the external density profile.\\

 In the direction of this burst, 17 events with photon energies greater than 10 GeV in the rest frame were reported, with the highest at 126.1 GeV and the lowest at 14.5 GeV. Both events arrived at 243 s and 23.2 s after the trigger time, respectively.   A derivation of the maximum energy of synchrotron radiation for radiative/injection regime and with/without energy displays these photons can be hardly explained in the synchrotron afterglow scenario (see Eqs. \ref{ene_max_eps} and \ref{ene_max_q}). Even while the temporal and spectral indices of GRBs recorded by Fermi-LAT may be understood in the synchrotron scenario, it cannot explain photons with energy exceeding the synchrotron limit for any value of $\epsilon$ and $q$.  If the magnetic field decays downstream of the shock on a sufficiently small scale, the maximum synchrotron frequency may be increased by a factor of a few over the one-zone limit \citep{2012MNRAS.427L..40K}.  In consideration of the above, a new mechanism, such as SSC or hadronic interactions, is most suggested rather than the synchrotron scenario.  Potentially high-energy photons exceeding the synchrotron limit may be produced through photo-hadronic interactions \citep[e.g., see][]{1997PhRvL..78.2292W}, but the IceCube collaboration's reports of no neutrino-GRB coincidences put a severe limit on how many hadrons may be in play during these collisions \citep{2012Natur.484..351A, 2016ApJ...824..115A, 2015ApJ...805L...5A}. 

\begin{figure*} 
\centering
\begin{tabular}{ccc}
    \includegraphics[width=0.5\textwidth]{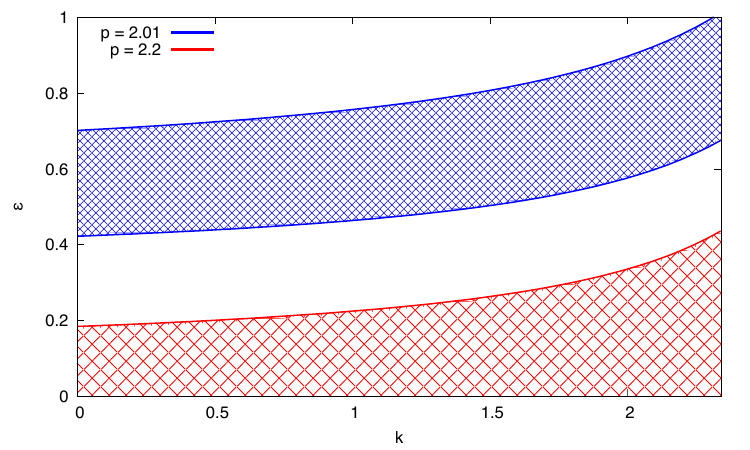}
    \includegraphics[width=0.5\textwidth]{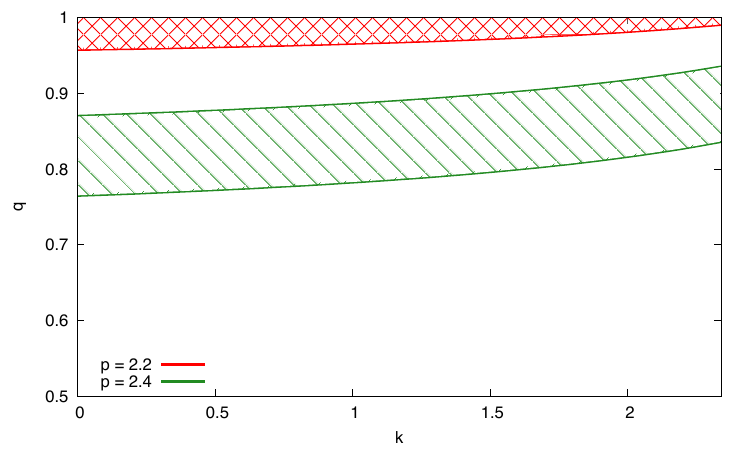} \\
\end{tabular}
\caption{The allowed parameter spaces that satisfy the temporal and spectral evolution of the LAT light curve of GRB 130427A.    The left-hand panel shows the allowed values of $k$ and $\epsilon$ for $p=2.01$ and $2.2$ and the right-hand panel displays the allowed values of $k$ and $q$ for $p=2.2$ and $2.4$.} \label{Fig3}
\end{figure*}

Considering the CRs reported in Tables \ref{Table1} and \ref{Table2}, we plot the set of parameters $k$, $\epsilon$ and $q$ that satisfy the temporal and spectral indexes reported in \cite{Ajello_2019} for GRB 130427A. The left-hand panel in Figure \ref{Fig3} shows the allowed parameter space  of $k$ and $\epsilon$ for $p=2.01$ and $2.2$ and the right-hand panel displays the allowed values of $k$ and $q$ for $p=2.2$ and $p=2.4$.  These panels show that as $p$ increases $\epsilon$ decreases,  and as $p$ increases $q$ increases. Any value of $p>2.4$ ($p<2.2$) and $\epsilon>0$ ($q > 1$) satisfies the CRs.  It is worth noting that from this figure can be inferred that for the adiabatic regime ($\epsilon=0$) and without energy injection $q=1$, a spectral index of $p\simeq2.2$ is obtained, which agree the value reported in \cite{2013ApJ...779L...1K}. We want to highlight that a millisecond magnetar is discarded as progenitor.

It is worth noting that to explain the afterglow of GRB 130427A, \cite{2022MNRAS.515..555B} derived a set of new CRs and proposed that the emission results from an underlying jet with a shallow angular profile in which larger viewing angles gradually dominate the emission.

\subsection{Evolution of the magnetic microphysical parameter}

We can estimate the microphysical parameter associated with the total energy given to amplify the magnetic field using Eqs. \ref{break_ssc_hom} and \ref{syn_esp_win}, and the passage of the synchrotron cooling break through the Fermi-LAT band ($\nu^{\rm ssc}_{\rm c}=\nu_{\rm LAT}$). In the radiative and energy injection scenario, this parameter corresponds to the values of

{\small
\begin{equation}
\label{eps_B1}
\varepsilon_B \lesssim
2.5\times10^{-4}\left(\frac{1+z}{1.022}\right)^{-\frac{6(2+\epsilon)}{7(8-\epsilon)}}\, (1+Y(\gamma_c))^{-\frac87}  A_{k}^{\frac{7\epsilon-36}{7(8-\epsilon)}}E_{53}^{-\frac{20}{7(8-\epsilon)}}\Gamma_{0,2.8}^{\frac{20\epsilon}{7(8-\epsilon)}}t_{3}^{\frac{4(2\epsilon-1)}{7(8-\epsilon)}} \left(\frac{h\nu^{\rm ssc}_c}{100\,{\rm MeV}}\right)^{-\frac27}
\end{equation}
}

and

{\small
\begin{equation}
\label{eps_B2}
\varepsilon_B \lesssim
2.2\times10^{-4}\left(\frac{1+z}{1.022}\right)^{-\frac{3}{14}}\, (1+Y(\gamma_c))^{-\frac87}  A_{k}^{-\frac{9}{14}}E_{53}^{-\frac{5}{14}}t_{3}^{\frac{5q-6}{14}} \left(\frac{h\nu^{\rm ssc}_c}{100\,{\rm MeV}}\right)^{-\frac27} 
\end{equation}
}
respectively. The values of this parameter are in the range of values ($10^{-5}\leq\varepsilon_B\leq 10^{-1}$) reported by several authors after describing the multiwavelength afterglow observations
\citep{1999ApJ...523..177W, 2002ApJ...571..779P, 2003ApJ...597..459Y, 2005MNRAS.362..921P, 2013ApJ...772..152W, 2014ApJ...785...29S, 2015MNRAS.454.1073B}.

We can notice that for $Y(\gamma_c)\ll 1$, Eqs. \ref{eps_B1} and \ref{eps_B2} are not altered, and  for $Y(\gamma_c)\gg 1$, $\epsilon_B$ increases. For instance,  for  $Y(\gamma_c)\gg 1$ and with the spectral breaks in the hierachy $ \nu^{\rm syn}_{\rm m} <  h\nu^{\rm syn}_{\rm KN}(\gamma_*) < \nu^{\rm syn}_{\rm c} < \nu^{\rm syn}_{\rm KN}(\gamma_{\rm c})$, the new Compton parameter becomes $Y(\gamma_*)\simeq 1$ for typical values of GRB afterglow \citep[see ][]{2010ApJ...712.1232W}. In all cases with the chosen parameters, $\varepsilon_B$ lies in the range of values reported in \cite{2014ApJ...785...29S} when the SSC afterglow model is considered. It is essential to mention that with a different set of parameters, e.g., earlier times ($t\sim 1\,{\rm s}$), the spectral breaks change in the hierarchy, and therefore, the KN effect is included.   For instance, \cite{2010ApJ...712.1232W} discussed the most general range of parameters: $10^{-6}<\varepsilon_B < 10^{-1}$,  $10^{-3}< \frac{A_k}{\rm cm^{-3}} <10$ (with $k=0$) for $\varepsilon_e=0.1$ and $E=10^{54}\,{\rm erg}$, for $t=1\,{\rm s}$ and $t=10\,{\rm s}$ in a constant-density medium. They showed regions in their Fig. 1  where the spectral breaks move to other positions in the hierarchy, and therefore, the KN effect is relevant for $t=1\,{\rm s}$ and not for $t=10\,{\rm s}$, as discussed.

 \cite{2019ApJ...883..134T} methodically examined the CRs of 59 LAT-detected bursts, using the temporal and spectral indices. The authors showed that these bursts satisfy the CRs of the slow-cooling regime, but only with the condition of having an exceptionally small value of $\epsilon_B<10^{-7}$.  We found that in the SSC afterglow scenario, with or without energy injection,  this parameter lies in the range of $3.5\times10^{-5}\leq\epsilon_B\leq0.33$, similar to reported in other afterglow modelling \citep[e.g., see][]{2014ApJ...785...29S}.

\section{Summary}
\label{sec5}

We have extended the work presented in \cite{2022ApJ...934..188F} and derived the CRs of the SSC afterglow model evolving in a density-profile described by $n(r) \propto r^{\rm -k}$ with and without energy injection. Here, we have considered the values of density-profile index of $k=0.5$, $1$, $1.5$ and $2.5$,  and relativistic electrons decelerated during the adiabatic and radiative afterglow phase with spectral index of $1<p<2$ and $p>2$. We have performed an analysis between these CRs and the spectral and temporal indices of bursts that were reported in 2FLGC.  In the radiativo scenario without energy injection, we have summarized the number and percentage of bursts following each cooling condition.  This best-case scenario is most consistent with ${\rm max \{\nu^{\rm ssc}_{\rm m}, \nu^{\rm ssc}_{\rm c}\}<\nu}$ for $k=0.5$ (22 GRBs, 25.88\%), $k=1.0$ (22 GRBs, 25.88\%), $k=1.5$ (21 GRBs, 24.71\%) and $k=2.5$ (19 GRBs, 22.35\%).  There is one burst that satisfies the CRs for each value of $k$ when the SSC model evolve in the cooling condition  ${\rm \nu^{\rm ssc}_{\rm c}\leq \nu \leq  \nu^{\rm ssc}_{\rm m}}$, and few bursts for  ${\rm \nu^{\rm ssc}_{\rm m}\leq \nu \leq  \nu^{\rm ssc}_{\rm c}}$ for $k=0.5$ (3 GRBs, 3.53\%), $1$ (2 GRBs, 2.35\%) and $1.5$ (1 GRB, 1.18\%). In the adiabatic scenario without energy injection, we have summarized the number and percentage of bursts following each cooling condition.  The best-case scenario is most consistent with ${\rm max \{\nu^{\rm ssc}_{\rm m}, \nu^{\rm ssc}_{\rm c}\}<\nu}$ for $k=0.5$ (32 GRBs, 37.65\%), $k=1.0$ (32 GRBs, 37.65\%), $k=1.5$ (34 GRBs, 40.0\%) and $k=2.5$ (40 GRBs, 47.06\%).  For the cooling condition ${\rm \nu^{\rm ssc}_{\rm m}\leq \nu \leq  \nu^{\rm ssc}_{\rm c}}$, 28 (25 and 7) GRBs for $k=0.5$ ($1$ and $1.5$) fulfill the CRs.  There is one burst that satisfies the CRs for each value of $k$ when the SSC model evolve in the cooling condition  ${\rm \nu^{\rm ssc}_{\rm c}\leq \nu \leq  \nu^{\rm ssc}_{\rm m}}$.

The standard synchrotron forward-shock model, which is valid up to the synchrotron limit, is able to provide an adequate explanation for both the afterglow of a GRB as well as the values of its spectral and temporal indices. Both of these phenomena can be shown to be adequately described by the model. Although, we have analyzed GRB 130427A, there have been 29 bursts in 2FLGC that have reported photons with energies more than 10 GeV, which cannot be well defined by the synchrotron scenario.  We propose that the SSC afterglow model is more suitable for comprehending this sample of bursts than the synchrotron model, which predicts that the greatest photon energy released during the deceleration phase would be roughly $\approx 1 \, {\rm GeV}$.  Photo-hadronic interactions are ruled out since neutrino non-coincidences with GRBs have been recorded by the IceCube collaboration \citep{2012Natur.484..351A, 2016ApJ...824..115A, 2015ApJ...805L...5A}. This sample of 29 bursts may be best explained by SSC process, although the CRs of the synchrotron standard model in some cases could describe them.\\

After selecting 59 LAT-detected bursts, \cite{2019ApJ...883..134T} methodically examined their CRs using temporal and spectral indices. For example, they found that although the standard synchrotron emission describes the spectrum and temporal indices in most instances, there is still a large fraction of bursts that can poorly be described with this model.  They showed, however, that many GRBs satisfy the CRs of the slow-cooling regime, but only when the magnetic microphysical parameter has an exceptionally small value of $\epsilon_B<10^{-7}$.   Here, we show that the SSC afterglow model, with or without energy injection, produces a value of $3.5\times10^{-5}\leq\epsilon_B\leq0.33$ \citep[e.g., see][]{2014ApJ...785...29S}. Therefore, the CRs of SSC afterglow models are needed, together with the research provided by \cite{2019ApJ...883..134T}, to account for the bursts in the 2FLGC that cannot be explained in the synchrotron scenario (e.g., those with photon energies above 10 GeV).

\section*{Acknowledgements}

We thank Peter Veres and Tanmoy Laskar for useful discussions.  NF acknowledges financial support  from UNAM-DGAPA-PAPIIT  through  grant IN106521. 

\section*{Data Availability}

The data used by this article was taken from the Second Gamma-ray Burst Catalog \cite[2FLGC;][]{Ajello_2019}.





\bibliographystyle{mnras}
\bibliography{mnras_template} 

\newpage




\begin{table}[h]
\centering \renewcommand{\arraystretch}{1.85}\addtolength{\tabcolsep}{1pt}
\caption{CRs of the SSC forward-shock scenario evolving in a stratified environment and adiabatic/radiative regime. CRs are shown for fast- and slow-cooling regime without energy injection}
\label{Table1}
\begin{tabular}{ccccccc}
\hline
 &  & $\beta$ & $\alpha(p)$ & $\alpha(p)$ & $\alpha(\beta)$ & $\alpha(\beta)$ \\
 &  & & $(1<p<2)$ & $(p>2)$ & $(1<p<2)$ & $(p > 2)$ \\
\hline
&  &  & Slow cooling &  &  &  \\
\hline
1 & \scriptsize{$\nu^{\rm ssc}_m<\nu<\nu^{\rm ssc}_c$} & \scriptsize{$\frac{p-1}{2}$} & \scriptsize{$\scriptsize{\frac{4(7+4\epsilon)+k[6-p(2-\epsilon)-13\epsilon]}{4[2(4-k)-\epsilon]}}$} & \scriptsize{$-\frac{4(11-9p-4\epsilon)-k[(22-9\epsilon-p(10+\epsilon)]}{4[2(4-k)-\epsilon]}$} & \scriptsize{$\frac{2(7+4\epsilon)+k[2-6\epsilon-\beta(2-\epsilon)]}{2[2(4-k)-\epsilon]}$} & \scriptsize{$-\frac{4(1-2\epsilon-9\beta)-k[6-5\epsilon-\beta(10+\epsilon)]}{2[2(4-k)-\epsilon]}$} \\
2 & \scriptsize{$\nu^{\rm ssc}_c<\nu$} & \scriptsize{$\frac{p}{2}$} & \scriptsize{$\frac{8(4+\epsilon)-k[4+p(2-\epsilon)+6\epsilon]}{4[2(4-k)-\epsilon]}$} & \scriptsize{$-\frac{40-36p-2k(6-\epsilon)-8\epsilon+kp(10+\epsilon)}{4[2(4-k)-\epsilon]}$} & \scriptsize{$\frac{4(4+\epsilon)-k[2+\beta(2-\epsilon)+3\epsilon]}{2[2(4-k)-\epsilon]}$} & \scriptsize{$-\frac{20-36\beta-k(6-\epsilon)-4\epsilon+\beta k(10+\epsilon)}{2[2(4-k)-\epsilon]}$} \\
\hline
 &  &  & Fast cooling &  &  &  \\
\hline
3 & \scriptsize{$\nu^{\rm ssc}_c<\nu<\nu^{\rm ssc}_m$} & \scriptsize{$\frac{1}{2}$} & \scriptsize{$-\frac{4-k(2-3\epsilon)-8\epsilon}{4[2(4-k)-\epsilon]}$} & \scriptsize{$-\frac{4-k(2-3\epsilon)-8\epsilon}{4[2(4-k)-\epsilon]}$} & \scriptsize{$-\frac{[4-k(2-3\epsilon)-8\epsilon]\beta}{2[2(4-k)-\epsilon]}$} & \scriptsize{$-\frac{[4-k(2-3\epsilon)-8\epsilon]\beta}{2[2(4-k)-\epsilon]}$} \\
4 & \scriptsize{$\nu^{\rm ssc}_m<\nu$} & \scriptsize{$\frac{p}{2}$} & \scriptsize{$\frac{8(4+\epsilon)-k[4+p(2-\epsilon)+6\epsilon]}{4[2(4-k)-\epsilon]}$} & \scriptsize{$-\frac{40-36p-2k(6-\epsilon)-8\epsilon+kp(10+\epsilon)}{4[2(4-k)-\epsilon]}$} & \scriptsize{$\frac{4(4+\epsilon)-k[2+\beta(2-\epsilon)+3\epsilon]}{2[2(4-k)-\epsilon]}$} & \scriptsize{$-\frac{20-36\beta-k(6-\epsilon)-4\epsilon+\beta k(10+\epsilon)}{2[2(4-k)-\epsilon]}$} \\
\hline
\end{tabular}
\end{table}

\begin{table}
\centering \renewcommand{\arraystretch}{1.85}\addtolength{\tabcolsep}{0pt}
\caption{CRs of the SSC forward-shock scenario with energy injection evolving in a stratified environment. CRs are shown for fast- and slow-cooling regime.}
\label{Table2}
\begin{tabular}{ccccccc}
\hline
 &  & $\beta$ & $\alpha(p)$ & $\alpha(p)$ & $\alpha(\beta)$ & $\alpha(\beta)$ \\
 &  & & $(1<p<2)$ & $ (p>2)$ & $(1<p<2)$ & $ (p>2)$ \\
\hline
&  &  & Slow cooling &  &  &  \\
\hline

1 & \scriptsize{$\nu^{\rm ssc}_m<\nu<\nu^{\rm ssc}_c$} & \scriptsize{$\frac{p-1}{2}$} & \scriptsize{$-\frac{12-k[16-p(2-q)-13q]-26q}{4(4-k)}$} & \scriptsize{$-\frac{36-14q-6p(2+q)-k[20-9q-p(4+q)]}{4(4-k)}$} & \scriptsize{$-\frac{6-k[7-6q-\beta(2-q)]-13q}{2(4-k)}$} & \scriptsize{$-\frac{12-10q-6\beta(2+q)-k[8-5q-\beta(4+q)]}{2(4-k)}$} \\
2 & \scriptsize{$\nu^{\rm ssc}_c<\nu$} & \scriptsize{$\frac{p}{2}$} & \scriptsize{$\frac{16q+k[4-p(2-q)-6q]}{4(4-k)}$} & \scriptsize{$-\frac{4(6-q)-6p(2+q)-k[2(4-q)-p(4+q)]}{4(4-k)}$} & \scriptsize{$\frac{8q+k[2-\beta(2-q)-3q]}{2(4-k)}$} & \scriptsize{$-\frac{2(6-q)-6\beta(2+q)-k[4-q-\beta(4+q)]}{2(4-k)}$} \\
\hline
 &  &  & Fast cooling &  &  &  \\
\hline
3 & \scriptsize{$\nu^{\rm ssc}_c<\nu<\nu^{\rm ssc}_m$} & \scriptsize{$\frac{1}{2}$} & \scriptsize{$-\frac{12-10q-4k+3kq}{4(4-k)}$} & \scriptsize{$-\frac{12-10q-k(4-3q)}{4(4-k)}$} & \scriptsize{$-\frac{[12-10q-4k+3kq]\beta}{2(4-k)}$} & \scriptsize{$-\frac{[12-10q-4k+3kq]\beta}{2(4-k)}$} \\
4 & \scriptsize{$\nu^{\rm ssc}_m<\nu$} & \scriptsize{$\frac{p}{2}$} & \scriptsize{$\frac{16q+k[4-p(2-q)-6q]}{4(4-k)}$} & \scriptsize{$-\frac{4(6-q)-6p(2+q)-k[2(4-q)-p(4+q)]}{4(4-k)}$} & \scriptsize{$\frac{8q+k[2-\beta(2-q)-3q]}{2(4-k)}$} & \scriptsize{$-\frac{2(6-q)-6\beta(2+q)-k[4-q-\beta(4+q)]}{2(4-k)}$} \\
\hline

\end{tabular}
\end{table}

\begin{table}
    \centering\renewcommand{\arraystretch}{1.4}
    \caption{Number and percentage of bursts satisfying each CR, summarized for the SSC forward-shock model in partially radiative regime ($\epsilon=0.5$) without energy injection.}
    \label{Table3}
    \begin{tabular}{c c c c c c}
     \hline
     \hline
    
      $\nu$ Range & $k$  & CR: 1 $<$ p $<$ 2  & CR:  2 $<$ p & GRBs Satisfying Relation & Proportion Satisfying Relation \\
     \hline
     
    ${\rm \nu_m^{\rm ssc} < \nu < \nu_{c}^{\rm ssc}}$ & $0.5$ &  $\frac{10(3+\epsilon)-\beta(2-\epsilon)}{4(7-\epsilon)}$ & $-\frac{2-13\epsilon-\beta(62-\epsilon)}{4(7-\epsilon)}$ & 3 & 3.53\% \\\cline{2-6}
     & 1.0&  $\frac{2(8+\epsilon)-\beta(2-\epsilon)}{2(6-\epsilon)}$ & $\frac{2+3\epsilon+\beta(26-\epsilon)}{2(6-\epsilon)}$  & 2 & 2.35\% \\\cline{2-6}
     & 1.5 &   $\frac{2(17-\epsilon)-3\beta(2-\epsilon)}{4(5-\epsilon)}$  & $\frac{10+3\epsilon+3\beta(14-\epsilon)}{4(5-\epsilon)}$  & 1 & 1.18\% \\\cline{2-6}
     & 2.5&   $\frac{2(19-7\epsilon)-5\beta(2-\epsilon)}{4(3-\epsilon)}$ & $\frac{22-9\epsilon +\beta(22-5\epsilon)}{4(3-\epsilon)}$  & 0 & 0.00\% \\
    \hline
    \hline
     ${\rm \nu_c^{\rm ssc} < \nu < \nu_{m}^{\rm ssc}}$ & 0.5 & $-\frac{(6-5\epsilon)\beta}{2(7-\epsilon)}$ & $-\frac{(6-5\epsilon)\beta}{2(7-\epsilon)}$ & 1 & 1.18\% \\\cline{2-6}
     & 1.0  & $-\frac{(2-5\epsilon)\beta}{2(6-\epsilon)}$  & $-\frac{(2-5\epsilon)\beta}{2(6-\epsilon)}$ & 1 & 1.18\% \\\cline{2-6}
     & 1.5  & $-\frac{(2+\epsilon)\beta}{2(5-\epsilon)}$  & $-\frac{(2+\epsilon)\beta}{2(5-\epsilon)}$ & 1 & 1.18\% \\\cline{2-6}
     &  2.5  & $-\frac{(7\epsilon-2)\beta}{2(3-\epsilon)}$ & $-\frac{(7\epsilon-2)\beta}{2(3-\epsilon)}$ & 1 & 1.18\% \\
     \hline
     \hline
     ${\rm max\{\nu_m^{\rm ssc}, \nu_c^{\rm ssc}\} < \nu}$ & 0.5  & $\frac{8(4+\epsilon)-[2+\beta(2-\epsilon)+3\epsilon]}{4(7-\epsilon)}$ & $-\frac{20-36\beta-\frac12(6-\epsilon)- 4\epsilon + \frac{\beta}{2}(10+\epsilon)}{2(7-\epsilon)}$ & 22 & 25.88\% \\\cline{2-6}
     & 1.0  & $\frac{4(4+\epsilon)-[2+\beta(2-\epsilon)+3\epsilon]}{2(6-\epsilon)}$  & $-\frac{20-36\beta-(6-\epsilon)- 4\epsilon + \beta(10+\epsilon)}{2(6-\epsilon)}$ & 22 & 25.88\% \\\cline{2-6}
     & 1.5  & $\frac{8(4+\epsilon)-3[2+\beta(2-\epsilon)+3\epsilon]}{4(5-\epsilon)}$  & $-\frac{20-36\beta-\frac32(6-\epsilon)- 4\epsilon + \frac{3\beta}{2}(10+\epsilon)}{2(5-\epsilon)}$ & 21 & 24.71\% \\\cline{2-6}
     &  2.5 & $\frac{8(4+\epsilon)-5[2+\beta(2-\epsilon)+3\epsilon]}{4(3-\epsilon)}$ & $-\frac{20-36\beta-\frac52(6-\epsilon)- 4\epsilon + \frac{5\beta}{2}(10+\epsilon)}{2(3-\epsilon)}$ & 19 & 22.35\% \\
     \hline
     \hline

    \end{tabular}

\end{table}

\begin{table}
    \centering\renewcommand{\arraystretch}{1.8}
    \caption{Number and percentage of bursts satisfying each CR, summarized for the SSC forward-shock model in adiabatic regime with energy injection ($q=0.5$).}
    \label{Table4}
    \begin{tabular}{c c c c c c}
     \hline
     \hline
    
      $\nu$ Range & $k$  & CR: 1 $<$ p $<$ 2  & CR:  2 $<$ p & GRBs Satisfying Relation & Proportion Satisfying Relation \\
     \hline
     
    ${\rm \nu_m^{\rm ssc} < \nu < \nu_{c}^{\rm ssc}}$ & $0.5$ &  $\frac{7q-5-\beta(2-q)}{7}$ & $-\frac{12 - 10q- 6\beta(2+q)-\frac12[8-5q-\beta(4+q)]}{7}$ & 28 & 32.94\% \\\cline{2-6}
     & 1.0&  $\frac{7q+1-\beta(2-q)}{6}$ & $-\frac{12 - 10q- 6\beta(2+q)-[8-5q-\beta(4+q)]}{6}$  & 25 & 29.41\% \\\cline{2-6}
     & 1.5 &   $\frac{9-5q-3\beta(2-q)}{5}$  & $-\frac{12 - 10q- 6\beta(2+q)-\frac32[8-5q-\beta(4+q)]}{5}$  & 7 & 8.24\% \\\cline{2-6}
     & 2.5&   $\frac{13-17q-5\beta(2-q)}{3}$ & $-\frac{12 - 10q- 6\beta(2+q)-\frac52[8-5q-\beta(4+q)]}{3}$  & 0 & 0.00\% \\
    \hline
    \hline
     ${\rm \nu_c^{\rm ssc} < \nu < \nu_{m}^{\rm ssc}}$ & 0.5 & $\frac{(17q-20)\beta}{14}$ & $\frac{(17q-20)\beta}{14}$ & 1 & 1.18\% \\\cline{2-6}
     & 1.0  & $\frac{(7q-8)\beta}{6}$  & $\frac{(7q-8)\beta}{6}$ & 1 & 1.18\% \\\cline{2-6}
     & 1.5  & $\frac{(11q-12)\beta}{10}$  & $\frac{(11q-12)\beta}{10}$ & 1 & 1.18\% \\\cline{2-6}
     &  2.5  & $\frac{(5q-4)\beta}{6}$ & $\frac{(5q-4)\beta}{6}$ & 1 & 1.18\% \\
     \hline
     \hline
     ${\rm max\{\nu_m^{\rm ssc}, \nu_c^{\rm ssc}\} < \nu}$ & 0.5  & $\frac{13q+2-\beta(2-q)}{14}$ & $-\frac{2(6-q) -  6\beta(2+q)-\frac12[4-q-\beta(4+q)]}{7}$ & 32 & 37.65\% \\\cline{2-6}
     & 1.0  & $\frac{5q+2-\beta(2-q)}{6}$  & $-\frac{2(6-q) -  6\beta(2+q)-[4-q-\beta(4+q)]}{6}$ & 32 & 37.65\% \\\cline{2-6}
     & 1.5  & $\frac{7q+6-3\beta(2-q)}{10}$  & $-\frac{2(6-q) -  6\beta(2+q)-\frac32[4-q-\beta(4+q)]}{5}$ & 34 & 40.00\% \\\cline{2-6}
     &  2.5 & $\frac{q+10-5\beta(2-q)}{6}$ & $-\frac{2(6-q) -  6\beta(2+q)-\frac52[4-q-\beta(4+q)]}{3}$ & 40 & 47.06\% \\
     \hline
     \hline

    \end{tabular}

\end{table}

\bsp	
\label{lastpage}
\end{document}